\documentclass[11pt]{article}

\usepackage{theorem,amssymb,amsbsy,latexsym}

\theoremstyle{plain}

\newcommand{\Wc}{\check W}

{\theorembodyfont{\upshape}}

\newcommand{\cC}{{\cal C}}

\newcommand{\cZ}{{\cal Z}} 
\newcommand{\cF}{{\cal F}} 
\newcommand{\xx}{\stackrel {\scriptscriptstyle \times}{\scriptscriptstyle \times}}

\newcommand{\tet}{\theta}
 
\newcommand{\id}{{\bf 1}}

\newcommand{\ee}[1]{{\rm e}^{#1}}
\newcommand{\ii}{{\rm i}}
\newcommand{\dd}{{\rm d}}
\newcommand{\f}{\frac}

\newcommand{\vx}{{\bf x}}
\newcommand{\vy}{{\bf y}}

\newcommand{\Ref}[1]{(\ref{#1})}
\newcommand{\eps}{\varepsilon}
\newcommand{\half}{\mbox{$\frac{1}{2}$}}
\newcommand{\third}{\mbox{$\frac{1}{3}$}}
\newcommand{\quarter}{\mbox{$\frac{1}{4}$}}

\newcommand{\C}{{\mathbb C}}
\newcommand{\Z}{{\mathbb Z}}

\newcommand{\cH}{{\cal H}}

\newcommand{\QED}{\hfill$\square$}

\newcommand{\eq}{\begin{equation}}
\newcommand{\eqend}{\end{equation}}
\newcommand{\eqa}{\begin{eqnarray}}
\newcommand{\nonueqa}{\begin{eqnarray*}}
\newcommand{\eqaend}{\end{eqnarray}}
\newcommand{\nonueqaend}{\end{eqnarray*}}
\newcommand{\nonu}{\nonumber \\ \nopagebreak}
\newcommand{\bma}[1]{\begin{array}{#1}}
\newcommand{\ema}{\end{array}}
\newcommand{\bc}{\begin{center}}
\newcommand{\ec}{\end{center}}

\newcounter{saveeqn}
\newcounter{App} 
\newcommand{\app}{%
\stepcounter{App}%
\setcounter{saveeqn}{\value{equation}}%
\setcounter{equation}{0}%
\renewcommand{\theequation}{\Alph{App}\arabic{equation}} }
\newcommand{\appende}{%
\setcounter{equation}{\value{saveeqn}}%
\renewcommand{\theequation}{\arabic{equation}}  }

\begin{document}
\begin{flushright}
July 12, 2005
\end{flushright}
\vspace{.4cm}

\begin{center}

{\Large \bf Remarkable identities related to the (quantum)
elliptic Calogero-Sutherland model}

\vspace{1 cm}

{\large Edwin Langmann}\\

\vspace{0.3 cm} {\em Mathematical Physics, Physics KTH, AlbaNova,
SE-106 91 Stockholm, Sweden}

\end{center}

\begin{abstract}
We present remarkable functional identities related to the elliptic
Calogero-Sutherland (eCS) system. We derive them from a second
quantization of the eCS model within a quantum field theory model of
anyons on a circle and at finite temperature. The identities involve
two eCS Hamiltonians with arbitrary and, in general, different
particle numbers $N$ and $M$, and a particular function of $N+M$
variables arising as anyon correlation function of $N$ particles and
$M$ anti-particles. In addition to identities obtained from anyons
with the same statistics parameter $\lambda$, we also obtain ``dual''
relations involving ``mixed'' correlation functions of anyons with two
different statistics parameters $\lambda$ and $1/\lambda$. We also
give alternative, elementary proofs of these identities by direct
computations.
\bigskip 

\noindent {PACS: 02.30.Ik, 03.65.-w\\ MSC-class: 35Q58, 81T40}
\end{abstract}

\section{Background and result} The elliptic Calogero-Sutherland
(eCS) system is a quantum mechanical model of an arbitrary number,
$N$, of particles moving on a circle of length $2\pi$ and interacting
with 2-body potentials given by the Weierstrass elliptic functions
$\wp$ \cite{C,Su,OP}. More specifically, this model is defined by the
differential operator
\eq
\label{eCS}
H = - \sum_{j=1}^N\frac{\partial^2}{\partial x_j^2} \; + \; 2 \lambda
(\lambda-1)\!\! \sum_{1\leq j<k\leq N} V(x_j-x_k) \eqend
where $-\pi\leq x_j\leq \pi$ are coordinates on the circle,
$\lambda>0$ is a real parameter determining the coupling strength, and
\eq
V(r) = \sum_{m\in\Z}\f{1}{ 4\sin^2[(r+ \ii\beta m )/2]} \: , \quad
\beta>0 
\eqend
is essentially the Weierstrass elliptic function $\wp$ with periods
$2\pi$ and $\ii \beta$,
\eq V(z) = \wp(z) + c_0 ,\quad c_0 = \frac{1}{12} -\sum_{m=1}^\infty
\frac{1}{2\sinh^2[(\beta m)/2]} \label{Vwp} \eqend
(see Appendix A.1. in \cite{EL4}, e.g.).

In this paper we obtain and prove various remarkable identities
involving eCS Hamiltonians and special functions of many variables
constructed from the following building block,
\eq 
\label{tet}
\tet(z) =  \sin(z/2) \prod_{n=1}^\infty
(1-2q^{2n}\cos(z) + q^{4n})\: , \quad q=\ee{-\beta/2} , 
\eqend 
which is essentially the Jacobi Theta function $\vartheta_1$,
\eqa
\label{Theta1} 
\tet(z) = \frac{1}{ 2 q^{1/4} \prod_{n=1}^\infty (1-q^{2n})}\;
\vartheta_1(\half z) \eqaend
(see Section 21.3 in \cite{WW}, e.g.). We derive these results from
the quantum field theory construction in \cite{EL2}.  We collect all
these identities in the following theorem.

\bigskip 

\noindent {\bf Theorem:} {\it Let
\eq
\label{FNM}
 F_{N,M}(\vx;\vy) = \f{ \prod_{1\leq j<k\leq N} \tet(x_{k}-x_j)^{\lambda} 
\prod_{1\leq j<k\leq M} 
\tet(y_{j}-y_{k})^{\lambda}}{\prod_{j=1}^N\prod_{k=1}^M
\tet(x_j-y_k)^{\lambda}} 
\eqend
with $\vx\in\C^N$ and $\vy\in\C^M$ and $\tet(z)$ defined in Eq.\
\Ref{tet}. This function obeys the identity
\eq
\label{rem2}
\left[H_{\lambda,N}(\vx) - H_{\lambda, M} (\vy) +
2(N-M)\lambda\frac{\partial}{\partial\beta} - c_{N,M} \right] F_{N,M} (\vx;\vy)
= 0 \eqend
with $H=H_{\lambda,N}(\vx)$ the eCS Hamiltonian in Eq.\ \Ref{eCS} and
the constant
\eqa c_{N,M} = \lambda^2 \Bigl[N(N-1)-M(M-1)\Bigr] c_0  \nonu +
(N-M)\lambda^2 \Bigl[ N(N-1) + M (M-1)-2 NM \Bigr]c_1 \label{cNM}
\eqaend
where 
\eq c_0 = \frac1{12} -\sum_{n=1}^\infty \frac{2
q^{2n}}{(1-q^{2n})^2},\quad c_1 = \frac1{12} . \label{c01} \eqend
Moreover, a similar identity holds true for the function
\eq
\label{tFNM}
\tilde F_{N,M}(\vx;\vy) = \prod_{1\leq j<k\leq N}
\tet(x_{k}-x_j)^{\lambda} \prod_{1\leq j<k\leq M}
\tet(y_{j}-y_{k})^{1/\lambda}\prod_{j=1}^N\prod_{k=1}^M
\tet(x_j-y_k) , \eqend
namely 
\eq
\label{trem2}
\left[H_{\lambda,N}(\vx) + \lambda H_{1/\lambda,M}(\vy) + 2(\lambda
N + M)\frac{\partial}{\partial\beta} - \tilde c_{N,M} \right] \tilde
F_{N,M} (\vx;\vy) = 0 \eqend
with 
\eqa \tilde c_{N,M} = \Bigl[ \lambda^2 N(N-1) + M(M-1)/\lambda + (1+\lambda)
NM\Bigr] c_0 \nonu + (\lambda N+M) \Bigl[\lambda N(N-1) + M(M-1)/\lambda + 2
NM\Bigr] c_1 . \label{tcNM} \eqaend
}
\bigskip

It is interesting to note that there are corresponding identities for
the first order differential operators
\eq
\label{PNX} 
P_N(\vx) =  \sum_{j=1}^N \ii \frac{\partial}{\partial x_j}  
\eqend
equal to the total momentum operator for the eCS system, namely
\eq
\left[P_N(\vx) +  P_M(\vy)\right]F_{N,M} (\vx;\vy) = 0 \label{later}
\eqend
(this is easily proven by observing that $F_{N,M} (\vx;\vy)$ is
invariant under common shifts $(x_j,y_k)\to (x_j+a,y_k+a)$ for
arbitrary real $a$), and similarly
\eq \left[P_N(\vx) + P_M(\vy)\right]\tilde F_{N,M} (\vx;\vy) = 0
\label{later1} .\eqend
It is natural to conjecture that similar identities hold true for all
commuting differential operators $H^{(j)}=H^{(j)}_{N}(\vx)$,
$j=1,2,,\ldots$, which are known to exist for the eCS model \cite{OP}
(since $H^{(j)}$ for $j=1$ and $2$ are equal to the total momentum
operator and the Hamiltonian of the eCS system, respectively).

\bigskip

\noindent {\em Remark 1.1:} Note that the constant $c_0$ in Eq.\ \Ref{c01}
is identical to the one in Eq.\ \Ref{Vwp}. Moreover, the constant
$c_1$ appears in the following from,
$$ 
c_1 = \frac1{8} -\sum_{n=1}^\infty \frac{n q^{2n}}{1-q^{2n}} -
 \frac{c_0}2, \label{c1} 
$$
but due to the identity 
\eq \sum_{n=1}^\infty \frac{n q^{2n}}{1-q^{2n}} = \sum_{m=1}^\infty
\frac{q^{2m}}{(1-q^{2m})^2} \label{identity} \eqend
one gets $c_1=1/12$.  \QED

\bigskip

\noindent {\em Remark 1.2:} It is interesting to note that by
redefining the elliptic functions by $\beta$-dependent constants as
follows,
\nonueqa V(r) &\to& V(r) + b_0\nonu \tet(r) &\to& B_1 \tet(r) \quad
\mbox{ with } \frac{\partial \log(B_1)}{\partial \beta} = b_1
\nonueqaend
the constants in our identities are changed as follows,
\nonueqa
c_{N,M} \to c_{N,M} -\lambda(\lambda-1) \Bigl[ N(N-1)-M(M-1) \Bigr]b_0 
\nonu - (N-M) \lambda^2 \Bigl[ N(N-1)+M(M-1)-2NM \Bigr]b_1   
\nonueqaend
and
\nonueqa \tilde c_{N,M} \to \tilde c_{N,M} - \Bigl[
\lambda(\lambda-1)N(N-1) + (1/\lambda -1) M(M-1) \Bigr]b_0 \nonu +
(\lambda N+M) \Bigl[ \lambda N(N-1) + M(M-1)/\lambda + 2NM\Bigr] b_1 .
\nonueqaend
In particular, choosing $b_0=c_0$ and $b_1=c_1$ we can simplify these
constants significantly,
\eqa c_{N,M} &\to& \lambda \Bigl[ N(N-1)-M(M-1) \Bigr]c_0 
\nonu 
\tilde c_{N,M}
&\to& \Bigl[\lambda N(N-1) + M(M-1) +(1+\lambda)MN \Bigr]c_0 .
\eqaend
This latter simple choice amounts to replacing
\eqa
V(r) &\to& \wp(r)\nonu 
\tet(r) &\to& \vartheta(\half r) . 
\eqaend
This shows that if one uses the standard elliptic functions one gets
somewhat simpler formulas. However, our choice has the advantage that
the trigonometric limit $q=0$ is not singular. Moreover, it removes a
trivial contribution from the eigenvalues of the eCS model
\cite{EL4}. \QED

\bigskip 

It is useful to write the constants above as follows,
\eq
c_{N,M} = \frac1{12}\lambda^2 (N-M) [(N-M)^2-1] + 2\lambda^2(N-M)(N+M-1) c_2 
\eqend
and
\eqa \tilde c_{N,M} = \frac1{12} [ \lambda^2 N^3 + M^3/\lambda +
3NM(\lambda N+M) - (\lambda^2 N+M/\lambda) ]\nonu - 
2[ \lambda^2 N^2 +
M^2/\lambda -(\lambda^2 N+ M/\lambda) + (\lambda+1)MN] c_2 \eqaend
with
\eq
c_2 = \sum_{n=1}^\infty \frac{n q^{2n}}{1-q^{2n}} . \label{c2} 
\eqend

In \cite{EL1,EL2} we obtained the special case $N=M$ of the identity
in Eq.\ \Ref{rem2} using a second quantization of the eCS model in a
quantum field theory (QFT) of anyons on a circle and at finite
temperature $1/\beta$, and in \cite{EL4}, Appendix A.3, we gave an
alternative, elementary proof by straightforward but rather tedious
computations. In this paper we show that the QFT results in \cite{EL2}
naturally imply all the identities in the Theorem above. To be
convincing also for readers not familiar with QFT we will also give
elementary proofs by direct computations which are, however, not so
illuminating. These direct proofs are based three functional
identities of the functions $V(r)$ and $\tet(r)$ introduced above:
Firstly,
\eq
\label{rel1}
V(r) \, = \, -\f{d^2}{d r^2} \log \tet(r) , 
\eqend
secondly, 
\eq
\label{rel2}
\phi(x)\phi(y) +\phi(x)\phi(z)+ \phi(y)\phi(z) 
= f(x)+f(y)+f(z)
\quad \mbox{ if $x+y+z=0$} 
\eqend
where 
\eq
\label{rel21}
\phi(x) = \f{d}{dx}\log\tet(x)\, , 
\quad f(x) = \frac12[V(x)-\phi(x)^2 - c_0] , \eqend
and thirdly, 
\eq
\label{rel3}
f(x) = -\frac{\partial}{\partial\beta}\log\theta(x) + c_1 
\eqend
with the constants $c_0$ and $c_1$ in Eq.\ \Ref{c01}. The (elementary)
proofs \Ref{rel1} and \Ref{rel2} can be found in \cite{EL4}, Appendix
A.1 and A.2, respectively; \Ref{rel3} follows readily from
$$\vartheta_1(x/2) = \sum_{n=1}^\infty (-1)^{n-1} q^{(n-1/2)^2}
\sin[(n-1/2) x]$$ (see Section 21.22 in \cite{WW}, e.g.) and Eq.\
\Ref{Theta1}, using the observation made in Remark 1.1 above. To avoid
misunderstanding we stress that the functional identities needed to
prove our results are classical and known since a long time.

An interesting special case of our result is for $M=0$ in which case
the identities reduce to
\eq H_{\lambda,N}(\vx) \Psi_0(\vx) = \left[
c_{N,0} - 2N\lambda\frac{\partial}{\partial\beta}\right]\Psi_0(\vx) \eqend
where $\Psi_0(\vx)=F_{N,0}(\vx,\vy)$, i.e.,
\eq \Psi_0(\vx) = \prod_{1\leq j<k\leq N} \tet(x_{k}-x_j)^{\lambda} .
\eqend
In the limiting case $\beta\to\infty$ the $\beta$-derivative term
disappears, and we recover the well-known eigenvalue equation for the
groundstate of the Sutherland model. It is interesting to note that
the elliptic generalization of this identity which we find here does
{\em not} give the groundstate of the eCS model, which is why
Sutherland's solution \cite{Su} of the Sutherland model does not
generalize to the elliptic case. For $N=2$, $M=0$ our identities
reduce to
$$ \left[-2\frac{\partial^2}{\partial x^2} + 4\lambda
\frac{\partial}{\partial\beta} -c_{2,0}\right]\theta(x)^\lambda=0,
$$ which for $\lambda=1$ is (essentially) the heat equation obeyed by
$\vartheta_1$ ($x=x_1-x_2$). Other interesting special cases will be
discussed in Section~3. 

It is worth noting that original motivation to study the QFT model of
anyons in \cite{CL} was its relation to the fractional quantum Hall
effect \cite{Wen}, and it is remarkable that this physics relation
proved to be helpful for finding the generalization in
\cite{EL1,EL2}. However, one can regard this construction also
pragmatically as a useful generating function technique for deriving
interesting identities which, once found, can also be proven by direct
computations. Since we believe that there are other identities to be
found along similar lines, we hope it is nevertheless of some interest
to not only give the direct proof but also the detailed QFT
derivation.

We also mention that the QFT construction in \cite{EL2} seems closely
related to earlier work on conformal field theory on the torus
\cite{B,EK,EFK,FW1,FW2} also finding an exploiting interesting
relations to the eCS model. The approach in \cite{EL2} and here is,
however, rather different in spirit and technique.

The plan of this paper is as follows. In Section~2 we give an outline
of how to derive our identities from the QFT results in
\cite{EL1,EL2}, emphasizing the physical interpretation and deferring
computational details to Appendix~A. In this discussion we clarify
this QFT construction, in particular the interpretation of the
representation we use as finite temperature representation (which,
since not needed there, was only discussed in an Appendix in
\cite{EL2}). Section~3 contains our conclusions, including a
comparison with previous results and a short discussion of possible
applications of our identities. The elementary, alternative proofs can
be found in Appendix~B.

\section{Quantum field theory derivation.}
In this Section we explain how the identities in our Theorem are
obtained from our results in \cite{EL1,EL2}. Some computational
details are deferred to Appendix~A.

\bigskip

\noindent {\bf Second quantization of the eCS model.} We first briefly
summarize the construction of anyons \cite{EL2}.

Anyons (for us) are operator valued distributions $\phi_\nu(x)$
parameterized by a coordinate on the unit circle, $-\pi\leq x\leq \pi$,
and depending on a real parameter $\nu$ determining their commutator
relations as follows,
\eq \phi_\nu(x)\phi_\nu(y) = \ee{\pm \ii \pi
\nu^2}\phi_\nu(y)\phi_\nu(x) \quad \mbox{ for $x\neq y$} \eqend
(see the Definition in Section~2.2 in \cite{EL2}; note that what we
denote as $\phi_\nu(x)$ here is identical with $\lim_{\eps\downarrow}
\phi_\eps^1(x)$ there; we ignore the regularization parameter $\eps$
here but indicate the parameter $\nu$ determining the statistics
instead: see also Remark~2.1 below). We constructed a particular
representation of these anyons on a Fock space $\cF$ generated from a
``vacuum'' $\Omega$ such that
\eq \langle \Omega,\phi_\nu(x)^* \phi_\nu(y) \Omega \rangle =
const.\, \theta(x)^{-\nu^2} \eqend
where $\theta(r)$ is the elliptic function in Eq.\ \Ref{tet}; $\langle
\cdot ,\cdot \rangle$ is the Hilbert space inner product and $*$ the
Hilbert space adjoint. This representation is characterized by an
parameter $\beta>0$ which determines the modulus of the elliptic
functions as $q=\exp(-\beta/2)$ and which, as we showed, has a natural
physical interpretation as inverse temperature.  We then constructed a
self-adjoint operator $\cH$ in $\cF$ which has remarkable commutator
relations with products of an arbitrary number $N$ of anyon fields
\eq \Phi^N_{\nu}(\vx) := \phi_\nu(x_1) \cdots \phi_\nu(x_N) ,
\label{0} \eqend
namely
\eq [\cH, \Phi^N_{\nu}(\vx)] \Omega = H_{N,\lambda}(\vx)
\Phi^N_{\nu}(\vx)\Omega
\label{1}
\eqend
where $H=H_{\lambda,N}(\vx)$ is the eCS Hamiltonian given in Eq.\
\Ref{eCS} with the coupling determined by the statistic parameter as
follows,
\eq \lambda = \nu^2 .  \eqend
These relations suggest to regard $\cH$ as a second quantization of
the eCS Hamiltonian. Another important property of $\cH$ is the
following,
\eq \langle \Omega, [A,\cH] \Omega \rangle = 0 , \label{2} \eqend
which is true for a large class of operators $A$ on $\cF$ (see Lemma 4
in \cite{EL2} for the precise statement), including arbitrary products
of anyons. 

\bigskip
\noindent {\em Remark 2.1:} A main technical point in \cite{EL2} was
to give precise mathematical meaning to this operator valued
distributions $\phi_\nu(x)$ by introducing approximate anyons
depending on a regularization parameter $\eps>0$ such that, for
$\eps>0$, they are well-defined operator, and the anyons are obtained
as a limit $\eps\downarrow 0$ (see Section~2.2 in \cite{EL2}). This is
a useful method to treat QFT divergences. However, in the present
paper we can ignore this technicality except for one instance in
Appendix~A.
\bigskip

\noindent {\em Remark 2.2:} It is worth noting that, at zero
temperature, the ``vacuum'' is a highest weight state annihilated by
$\cH$, but this property is lost a finite temperature. The relation in
Eq.\ \Ref{2} is a weaker substitute for this highest weight conditions
which holds true also at finite temperature. A similar remark applies
to the operator $\Wc^2$ discussed below.
\bigskip

\noindent {\bf Identities: a special case.}  In \cite{EL1,EL2} we
observed that the relations above imply that the anyon correlation
function
\eq G_{N,N}(\vx,\vy) \equiv \langle \Omega, \Phi^N_\nu(\vx)^*
\Phi^N_\nu(\vy) \Omega \rangle \label{F} \eqend
satisfies the following remarkable identity,
\eq \Bigl[ H_{\lambda,N}(\vx) -H_{\lambda,N}(\vy)\Bigr] G_{N,N}(\vx,\vy)
= 0 . \label{id0} \eqend
The argument is simple: using Eq.\ \Ref{2} for $A=\Phi^N_\nu(\vx)^*
\Phi^N_\nu(\vy)$ we get
\eq \langle [\cH, \Phi^N_\nu(\vx) ]\Omega, \Phi^N_\nu(\vy)\Omega
\rangle - \langle  \Omega,\Phi^N_\nu(\vx)^* [\cH , \Phi^N_\nu(\vy) ]
\Omega \rangle =0 \eqend
where we used $[\Phi^N_\nu(\vx)^*,\cH] = [\cH,\Phi^N_\nu(\vx)]^*$
which holds true since $\cH$ is self-adjoint. Inserting Eq.\ \Ref{1}
twice, moving the eCS Hamiltonians in front of the Hilbert inner
product, and using \Ref{F} we get Eq.\ \Ref{id0}. Computing
$G_{N,N}(\vx,\vy)$ one finds that it is equal to $F_{N,N}(\vx,\vy)$ in
Eq.\ \Ref{FNM} (see Proposition~1 in \cite{EL2} or Eqs.\ \Ref{G1} and
\Ref{G2} below), and we thus obtain the special case $N=M$ of the
identity in Eq.\ \Ref{rem2}.

\bigskip

\noindent {\bf Other correlation functions.} It is natural to try to
generalize this identity by considering generalized anyon correlation
functions which are vacuum expectation values of
$$
\Phi^N_\nu(\vx)^* \Phi^M_{\mu}(\vy) , 
$$
allowing also for different particle numbers in $\vx$ and $\vy$ and/or
different anyon parameters $\nu$ and $\mu$. To construct and compute
non-trivial such functions we need to recall a few more details of the 
anyon construction in \cite{EL2}.

The explicit form of the anyons is,
\eq \phi_\nu(x) = \; \xx R \exp \left( -\ii \nu^2 Q x - \nu
\sum_{n\neq 0} \frac{1}{n} \rho(n) \ee{\ii nx} \right) \xx
\label{anyon} \eqend
where the $\rho(n)$, $n\in \Z$, and $R$ are generators of the 
Heisenberg algebra defined by the following
relations
\eq [\rho(n),\rho(m)] = n\delta_{n,-m} ,\quad [\rho(n),R] =
\delta_{n,0} R \label{H1} \eqend
and 
\eq \rho(n)^* = \rho(-n),\quad R^* = R^{-1}
\label{H2}   \eqend
where
\eq \rho(0) = Q   \eqend
has the physical interpretation of a {\em charge operator}.  A
important point here is the definition of normal ordering
$\xx\cdot\xx$ which is not the standard one; see Lemma 1 in \cite{EL2}
for a detailed characterization. The operator $Q$ satisfies, by
definition,
\eq Q\Omega = 0, \eqend
and this has important consequences: we say that an operator $A$ on
$\cF$ has charge $q$ iff $[Q,A]=qA$, and the definitions above imply
that only operators $A$ with charge zero can have a non-zero vacuum
expectation value, and that $R^{\pm 1}$ changes the charge by $\pm
1$. We therefore need to to insert an appropriate power of $R$ to get
a non-trivial vacuum expectation value, and the natural correlation
function to consider is
\eq
G_{\nu,\mu;N,M}(\vx; \vy) = \langle \Omega, \Phi^N_\nu(\vx)^* R^{N-M}
\Phi^M_{\mu}(\vy) \Omega \rangle . \label{G1}
\eqend
By straightforward computations we obtain (for details see
Appendix~A.1)
\eqa G_{\nu,\mu;N,M}(\vx; \vy) = \ee{\ii(p_1 X - p_2 Y)}\frac{ 
\prod_{1\leq j<k\leq N} \tet(x_j-x_k)^{\nu^2} \prod_{1\leq j<k\leq M}
\tet(y_k-y_j)^{\mu^2}}{ \prod_{j=1}^N \prod_{k=1}^M
\tet(x_j-y_k)^{\nu\mu}} \label{G2} \eqaend
where the exponential factor gives the dependence of the
center-of-mass coordinates, 
\eq
X=\sum_{j=1}^N x_j,\quad Y=\sum_{j=1}^M y_j ,\label{XY} 
\eqend
and the center-of-mass momenta are determined by the statistics
parameters of the anyons as follows,
\eq
p_1 = \half(\nu^2 N-\nu\mu M),\quad p_2 = \half(\mu^2 M-\nu\mu N).  
\eqend
Note that, up to the exponential factors, these correlation functions
for $\mu=\nu=\sqrt{\lambda}$ and $\nu=-1/\mu=\sqrt{\lambda}$ are equal
to the functions in Eq.\ \Ref{FNM} and \Ref{tFNM}, respectively.

In the following we give a simplified derivations of Eqs.\ \Ref{rem2}
and \Ref{trem2}, ignoring less interesting terms which only contribute
to the constants $c_{N,M}$ and $\tilde c_{N,M}$, respectively, and
which we indicate by dots. This simplifies the argument
considerably. The complete equations including all terms are given in
Appendix A.3 and A.4, respectively.

\bigskip

\noindent {\bf Derivation of the identity in Eq.\ \Ref{rem2}.}
We now consider the function
\eq G_{N,M}(\vx;\vy) \equiv G_{\nu,\nu; N,M}(\vx;\vy) ,\quad
\nu=\sqrt{\lambda} \label{GNM}   \eqend
and try to use a similar argument as above. If we now use the identity
in Eq.\ \Ref{2} for $A=\Phi^N_\nu(\vx)^* R^{N-M} \Phi^M_{\nu}(\vy)$
and use Eq.\ \Ref{1} twice we obtain
\eq \Bigl[ H_{\lambda,N}(\vx) -H_{\lambda,M}(\vy)\Bigr]
G_{N,M}(\vx,\vy) = \langle\Omega, 
\Phi^N_\nu(\vx)^*[\cH,R^{N-M}]\Phi^M_\nu(\vy) \Omega\rangle  \label{ab1} \eqend
where we now get an additional term. It is a rather surprising that it
is possible to compute this term in a simple manner: this is due to a
``miracle'' which we now describe.

For that we need to recall the explicit form of the 2nd quantized eCS
Hamiltonian $\cH$, 
\eq \cH = \nu \Wc^3 + (1-\nu^2) \cC + 2\nu^2 \Wc^2 Q +\third\nu^4 Q^3
-\nu^4 c_0 Q , \label{cH} 
\eqend
with the constant $c_0$ in Eq.\ \Ref{c01}, 
\eq \Wc^s = \frac1s \int_0^{2\pi} \frac{\dd x}{2\pi} \xx \left(
\sum_{n\neq 0} \rho(n) \ee{\ii nx} \right)^s \xx \quad \mbox{ for
$s=2,3$} , \eqend
and $\cC$ is some (known) operator satisfying
\eq [\cC,R]=0,\quad \cC\Omega =0  \eqend
(see Proposition~2 in \cite{EL2}; the formula given there looks
different but is equivalent, as is seen by a simple computation and 
the identity in Eq.\ \Ref{identity}). 

\bigskip
\noindent {\em Remark 2.3:} The charge operator $Q$ (equal to the zero
mode $\rho(0)$) plays an even more important role here than in
\cite{EL2}, and it is therefore more natural now to write $\cH$ in
terms of the operators $\Wc^s$ with all zero modes removed: since
these operators obviously commute with the charge-rising operator $R$,
\eqa
\Wc^2 &=& \frac12 \sum'_{n} \xx
\rho(n)\rho(-n)\xx\nonu
\Wc^3 &=& \third \sum_{n,m}' \xx
\rho(n)\rho(m)\rho(-n-m) \xx
\eqaend
with the prime on the sums indicating that all terms with at least one
factor $\rho(0)$ are left out, it thus becomes easier to compute the
commutator of $\cH$ with $R$. It is gratifying to see that this also
simplifies the formula for $\cH$ (compare Eq.\ \Ref{cH} above with
Eq.\ (57) in \cite{EL2}).

We thus get
\eq [\cH, R^{N-M}] = 2(N-M)\lambda R^{N-M} \Wc^2 +\ldots , \label{4} \eqend
with the dots less interesting terms which we ignore for simplicity
(they are given in Appendix~2.3). The ``miracle'' is that this
commutator is proportional to the operator $\Wc^2$ which plays an
important two-fold role in the anyon QFT: firstly, $\Wc^2$ is
(essentially) the second quantization of the total momentum operator,
\eq
\label{WW21}
[\Wc^2,\Phi^N_\nu(\vx) ] = P_N(\vx) \Phi^N_\nu(\vx) + \ldots
\eqend
with $P_N(\vx)$ defined in Eq.\ \Ref{PNX} is the generator of
translations (see Eq.\ (68) in \cite{EL2}), and secondly, $\Wc^2$ is
(essentially) identical with the many-body Hamiltonian used to
construct the temperature representation (see Proposition~4 in
Appendix B.2 of \cite{EL2}) which implies
\eq
\label{WW22}
\langle \Omega, A \Wc^2\Omega \rangle = -\frac{\partial}{\partial
\beta} \langle \Omega, A\Omega \rangle + \ldots . \eqend

\bigskip

\noindent {\em Remark 2.4:} Eq.\ \Ref{WW22} is the key to our
identities. While the first term on the r.h.s. is easy to understand
from what we said above, the computation of the terms `$\ldots$'
proportional to $\langle A \rangle $ is somewhat subtle. This
computation clarifies some interesting aspects of our finite
temperature representation not mentioned in \cite{EL2}. They are
discussed in Appendix A.2.

\bigskip

Using Eqs.\ \Ref{WW21} and \Ref{WW22} we can compute
\nonueqa
\langle\Omega, \Phi^N_\nu(\vx)^*[\cH,R^{N-M}]\Phi^M_\nu(\vy)
\Omega\rangle 
= 2\lambda (N-M) \nonu \times \langle\Omega, \Phi^N_\nu(\vx)^*R^{N-M} 
\Bigl( [\Wc^2 ,\Phi^M_\nu(\vy)] + \Phi^M_\nu(\vy) \Wc^2\Bigr) 
\Omega\rangle +\ldots \nonu 
= 2\lambda (N-M) \Bigl( P_M(\vy) - \frac{\partial}{\partial \beta}\Bigr) G_{N,M}(\vx;\vy) 
+ \ldots ,   
\nonueqaend
and inserting this in Eq.\ \Ref{ab1} we obtain 
\eqa \Bigl[ H_{\lambda,N}(\vx) -H_{\lambda,M}(\vy)\Bigr] G_{N,M}(\vx,\vy) =
2\lambda (N-M) \Bigl( P_M(\vy) - \frac{\partial}{\partial \beta}\Bigr)
G_{N,M}(\vx;\vy) + \ldots .  \label{id2} \eqaend
From Eqs.\ \Ref{GNM}, \Ref{G2} and \Ref{FNM} we see that
$G_{N,M}(\vx;\vy)$ is equal to $F_{N,M}(\vx;\vy)$ up to a phase factor
depending only on center-of-mass coordinates $X$ and $Y$, and using
\eqa
\ee{-\ii p X}H_N(\vx) \ee{\ii p X} = H_N(\vx) + Np^2 -  2 p P_N(\vx) \nonu
\ee{-\ii p X}P_N(\vx) \ee{\ii p X} = P_N(\vx) -N p  \label{conj} 
\eqaend
the identity in Eq.\ \Ref{id2} turns into a similar identity for
$F_{N,M}(\vx;\vy)$. Remarkably we can use Eq.\ \Ref{later} to cancel
all terms involving $P_N(\vx)$ and $P_M(\vy)$. This proves the
identity in Eq.\ \Ref{rem2} up to the precise value of the constant
$c_{N,M}$.

As mentioned, the missing details to also compute the constant
$c_{N,M}$ are given in Appendix~A.2 and A.3.

\bigskip

\noindent {\bf Derivation of the identity in Eq.\ \Ref{trem2}.}  In
general, the argument above does not work for correlation functions in
Eq.\ \Ref{G1} if the anyon parameters $\nu$ and $\mu$ are
different. However, there is one such case where it does work: from
the explicit formula for the second quantization of the eCS
Hamiltonian in Eq.\ \Ref{cH} we observe that replacing $\nu$ by
$-1/\nu$ gives back essentially the same operator up to a constant
factor,
\eq \cH^{(\nu)} = -\lambda\cH^{(-1/\nu)} + 2(\lambda+1) \Wc^2 Q +
(\lambda^2+1/\lambda)(\third Q^3- c_0 Q) , \quad \lambda=\nu^2,
\label{3} \eqend
where we now indicate also the anyon parameter $\nu$. This suggest that
we should be able to also get an identity for the correlation functions
\eq \tilde G_{N,M}(\vx;\vy) \equiv G_{\nu,-1/\nu;N,M}(\vx;\vy),\quad
\nu=\sqrt{\lambda} \label{tGNM} \eqend
defined in Eq.\ \Ref{G1}. Indeed, by a similar computation as above we
obtain
\nonueqa 0 = \langle \Omega, [ \Phi^N_\nu(\vx)^* R^{N-M}
\Phi^M_{-1/\nu}(\vy), \cH^{(\nu)} ] \Omega \rangle =
\Bigl[H_{\lambda,N}(\vx) + \lambda H_{1/\lambda,M}(\vy) \Bigr] \tilde
G_{N,M} \nonu - \langle \Omega,\Phi^N_{\nu}(\vx)^* \Bigl(
[\cH^{(\nu)}, R^{N-M}] - 2(\lambda+1) M R^{N-M} \Wc^2 \Bigr)
\Phi^M_{-1/\nu}(\vy)\Omega\rangle + \ldots \nonueqaend
where we used Eqs.\ \Ref{2} and \Ref{3} and twice Eq.\ \Ref{1}. As
mentioned, the dots indicate less interesting terms specified in
Appendix~A.4. Inserting Eqs.\ \Ref{4} and \Ref{WW21} we get, similarly
as above, 
\eqa \Bigl( H_{\lambda,N}(\vx) + \lambda
H_{1/\lambda,M}(\vy)\Bigr)\tilde G_{N,M}(\vx;\vy) \nonu =
\Bigl(2(N-M)P_M(\vy) - 2(\lambda N+M) \frac{\partial}{\partial\beta}
\Bigr) \tilde G_{N,M}(\vx;\vy) + \ldots . \eqaend
Since $\tilde G_{N,M}(\vx;\vy)$ is equal to $\tilde F_{N,M}(\vx;\vy)$
up to the center-of-mass phase factor (cf.\ Eqs.\ \Ref{tGNM}, \Ref{G2}
and \Ref{tFNM}) we can use Eq.\ \Ref{conj} to obtain a similar
identity for $\tilde F_{N,M}(\vx;\vy)$. Again all terms involving
$P_N(\vx)$ and $P_M(\vy)$ cancel due to Eq.\ \Ref{later1}, and we
obtain the identity in Eq.\ \Ref{trem2} up to the value of the
constant $\tilde c_{N,M}$.

The details of this computations can be found in Appendix A.4. 

\section{Conclusions} 
It is interesting to note that by introducing the operators
\eq
L_{N,\lambda}(\vx) = 2\lambda N \frac{\partial}{\partial\beta} + H_{N,\lambda}(\vx) 
\eqend
one can write the identities in Eqs.\ \Ref{rem2} and \Ref{trem2} as
follows,
\eq \Bigl(L_{\lambda,N}(\vx) - L_{M,\lambda}(\vy) - c_{N,M} \Bigr)
F_{N,M}(\vx;\vy) = 0 \label{r2} \eqend
and
\eq \Bigl(L_{\lambda,N}(\vx) + \lambda L_{1/\lambda,M}(\vy) - \tilde
c_{N,M} \Bigr) \tilde F_{N,M}(\vx;\vy) = 0 ,\label{tr2}
\eqend
respectively. The operator $L_N$ seems to be a special case of one
introduced by Bernard \cite{B}; see Eq.\ (6.1) {\em ff} in
\cite{EK}. The results in Ref.\ \cite{B} suggests that it should be
possible to interpret Eqs.\ (\ref{r2},\ref{tr2}) as Ward identities of
some conformal field theory on the torus; see \cite{FW1,FW2}. We also
note that the special case $M=0$ of Eq.\ \Ref{rem2} seems to be
identical with an identity given in the Remark after Theorem 4.1 in
Ref.\ \cite{FW2}.

We finally discuss possible applications of these identities. We first
discuss the trigonometric limit $\beta\to \infty$ in which case the
derivative terms proportional to $\partial / \partial\beta$ in Eqs.\
(\ref{rem2},\ref{trem2}) are absent. In particular, for $M=0$ we
recover the well-known eigenvalue equation for the ground state of the
Sutherland model which is the starting point of Sutherland's solution
of this model \cite{Su}, as already mentioned in Section~1. Another
interesting special case is the identity in Eq.\ \Ref{rem2} for $N=M$:
it seems to give an alternative construction of the Q-operator playing
a central role in \cite{KMS} deriving interesting explicit results for
the solution of the Sutherland system. Moreover, this very identity is
also the starting point of an alternative solution algorithm for the
Sutherland model \cite{EL3}.  Our generalization of this identity to
$M\neq N$ might allow to construct a generalized Q-operator relating
eigenfunctions of the Sutherland model for different particle numbers
and/or different coupling parameters $\lambda$ and $1/\lambda$. For
the general elliptic case $\beta<\infty$ the identity $N=M$ in Eq.\
\Ref{rem2} was used as a starting point for a perturbative algorithm
to solve the eCS model as a formal power series in $q^2$
\cite{EL4}. We speculate that it might be possible to also find
elliptic generalizations of the results in Ref.\ \cite{KMS} using our
identities (this is suggested to us by the interesting results on the
3-particle eCS system in Section~7 of Ref.\ \cite{Skl}).

\app
\section*{Appendix A: QFT derivation. Details.} 
In this Appendix we provide the details of the quantum field theory
derivation of the identities summarized in our Theorem.

\bigskip

\noindent {\bf A.1 Computation of the anyon correlation functions.}
Here we give more details of how to compute $G_{\nu,\mu;N,M}(\vx;\vy)$
defined in Eq.\ \Ref{G1} and obtain Eq.\ \Ref{G2}.

We know that
\eq \Phi^N_{\nu}(\vy) = b_N(\vx)^{\nu^2} \xx \Phi^N_{\nu}(\vx) \xx
,\quad b_N(\vx) = \prod_{1\leq j<k\leq N} \tet(x_j-x_k) \eqend
and 
\eqa \langle \Omega, \xx \Phi^N_{\nu}(\vx)^* \xx R^{N-M}
 \xx \Phi^M_{\mu}(\vy)\xx\Omega \rangle \nonu = \langle \Omega, \ee{\ii\nu^2
 X/2}R^{-N} \ee{\ii\nu^2 X/2} R^N R^{-M} \ee{-\ii\mu^2 Y/2}R^{M}
 \ee{-\ii\mu^2 Y/2} \Omega \rangle (*) \nonu = \ee{\ii (N\nu^2 X
 - M\mu^2 Y)/2} (*) \eqaend
(see Eqs.\ (50), (16) and (17) and in \cite{EL2}; we also used
$R^{-1}Q R = Q+\id$ following from Eqs.\ (11) in \cite{EL2}) and
$$
(*) = \prod_{j,k}\ee{\nu\mu C(x_j-y_k)} = \prod_{j,k}
\tet(x_j-x_k)^{-\nu\mu} \ee{-\ii\nu\mu(x_j-y_k)/2}= 
\ee{-\ii \nu\mu( M X-NY)/2 } \prod_{j,k}
\tet(x_j-x_k)^{-\nu\mu}
$$
the other contributions (see Eq.\ (44) {\em ff} in \cite{EL2}). By
simple computations this yields Eq.\ \Ref{G2}.

\bigskip

\noindent {\bf A.2 Finite temperature correlations functions.}  Eq.\
\Ref{WW22} is a crucial step in the QFT derivation of our identities. As
discussed in the main text, it is a consequence of $\beta$ being equal
to the inverse temperature and $\Wc^2$ being (essentially) the
Hamiltonian used to construct the finite temperature representation of
our QFT model. We now discuss this relation in more detail.

In \cite{EL2}, Appendix B.1, we proved that the vacuum expectation
value of (essentially) any operator $A$ is equal to its thermal
expectation as follows,
\eq \langle \Omega, A \Omega \rangle = \frac1{\cZ} \lim_{a\to \infty}
{\rm Tr} (\ee{-\beta H_0} A_0) \label{aaa} \eqend
where
\eq H_0 = a Q_0^2 + \sum_{n=1}^\infty \rho_0(n) \rho_0(-n) \label{H0}
\eqend
and the subscripts `0' are to indicate that these operators are in the
standard (= zero temperature) representation;
\eq\cZ = \prod_{n=1}^\infty \frac1{(1-q^{2n})},\quad q=\ee{-\beta/2}
\eqend
is the partition function (see Proposition~4 and Eq.\ (B16) in
\cite{EL2}). Remarkably,
\eq
\frac1{\cZ} \frac{\partial\cZ}{\partial\beta} = -c_2 
\eqend
with $c_2$ the constant in Eq.\ \Ref{c2}, and thus Eq.\ \Ref{aaa}
implies
\eq \langle \Omega, A H \Omega \rangle = -\Bigl(
\frac{\partial}{\partial\beta} - c_2\Bigr) \langle \Omega, A \Omega
\rangle . \label{AH} \eqend
with $H = aQ^2 + \sum_{n=1}^\infty \rho(-n)\rho(n)$. Since $H$ is
(essentially) equal to $aQ^2+\Wc^2$ this implies Eq.\ \Ref{WW22}.

However, there is a subtle to point we need to take into account to
make this relation precise: the normal ordering prescription used in
\cite{EL2} is $\beta$-dependent, and $aQ^2 + \Wc^2$ is therefore {\em
not} equal to $H$ (which is defined by zero temperature normal
ordering) but differs from it by a constant. This difference can be
computed as follows: by definition of the normal ordering in the
thermal state (see Eq.\ (25) in \cite{EL2}), $ \langle \Omega, \Wc^2
\Omega \rangle = 0$, whereas Eq.\ \Ref{AH} for $A=\id$ gives $ \langle
\Omega, H \Omega \rangle= c_2$. We conclude that $\Wc^2 = H-aQ^2
-c_2$, and thus
\eq \langle \Omega, A \Wc^2 \Omega \rangle = -
\frac{\partial}{\partial\beta}\langle \Omega, A \Omega  \rangle 
\label{AH1} \eqend 
with the normal ordering difference taking away precisely the constant
$c_2$ in Eq.\ \Ref{AH}.

It is also important to note that also the anyon operators are defined
using normal ordering. More explicitly, as explained in Remark~2.6
after Eq.\ (46) in \cite{EL2}, normal ordering of the anyon field
$\phi_\nu(x)$ amounts to a multiplication with the constant
$$ \Bigl[ (1-\ee{-2\eps}) \prod_{m=1}^\infty
(1-q^{2m}\ee{-2\eps})^2 \Bigr]^{-\nu^2/2} $$
in the limit $\eps\downarrow 0$ (see Eq.\ (A4) {\em ff} in
\cite{EL2}); since the divergent factor is $\beta$ independent, the
factor accounting for the difference in normal ordering has a finite
limit as $\eps\downarrow 0$ which is identical with
$\cZ^{\nu^2}$. Thus
\eqa \langle \Omega, \Phi^N_\nu(\vx)^* R^{N-M} \Phi^M_{\mu}(\vy)
\Omega \rangle = \cZ^{N\nu^2 +M\mu^2-1} \lim_{a\to \infty} \nonu
\times {\rm Tr} \left( \ee{-\beta H_0} \Phi^N_{\nu,0}(\vx)^* R_0^{N-M}
\Phi^M_{\mu,0}(\vy) \right) .  \eqaend
We conclude that
\eqa \langle \Omega,\Phi^N_\nu(\vx)^* R^{N-M} \Phi^M_{\mu}(\vy) \Wc^2
\Omega \rangle = - \left( \frac{\partial}{\partial\beta} +
(N\nu^2+M\mu^2) c_2 \right) G_{\nu,\mu;N,M}(\vx;\vy) \label{7} \eqaend
(we used Eq.\ \Ref{G1}) which is the equation we need.  

\bigskip

\noindent {\bf A.3 Detailed derivation of Eq.\ \Ref{rem2}.} In our
derivation of Eq.\ \Ref{rem2} in the main text we ignored terms
proportional to $G_{N,M}(\vx;\vy)$ (indicated by dots), to simplify
the argument. Here we give the full derivation, with all terms
included.

Using Eq.\ \Ref{cH} we obtain 
\eq [\cH, R^{N-M}] = (2\lambda \Wc^2 - c_0\lambda^2 )(N-M) R^{N-M} +
\third\lambda^2 [Q^3,R^{N-M}] , \eqend
implying
\eqa \langle\Omega, \Phi^N_\nu(\vx)^*[\cH,R^{N-M}]\Phi^M_\nu(\vy)
\Omega\rangle = 2\lambda (N-M) \langle\Omega, \Phi^N_\nu(\vx)^* \Wc^2
R^{N-M}\Phi^M_\nu(\vy) \Omega\rangle \nonu + \lambda^2 [\third(N^3-M^3) -c_0(N-M) 
]G_{N,M}(\vx,\vy) . \eqaend

We now use
\eq [\Wc^2 + \half \lambda Q^2,\Phi^N_{\nu}(\vx) ] =
P_N(\vx)\Phi^N_{\nu}(\vx) \eqend
with $P_N(\vx)$ defined in Eq.\ \Ref{PNX} (see Eqs.\ (68) in
\cite{EL2}; there is a typo in this latter formula: $(\nu-1)(\nu-3)$
should be replaced by $(\nu^2-1)$). Using that and \Ref{7} for
$\nu=\mu$ (recall Eq.\ \Ref{GNM}) we obtain
\eqa \langle\Omega, \Phi^N_\nu(\vx)^* \Wc^2 R^{N-M}\Phi^M_\nu(\vy)
 \Omega\rangle = \Bigl( P_M(\vy) - \half \lambda M^2 \nonu -
 \frac{\partial}{\partial\beta} - \lambda(N+M) c_2 \Bigr)
 G_{N,M}(\vx,\vy) .  \eqaend
Putting the equations above together we obtain
\eqa \Bigl[ H_{\lambda,N}(\vx) -H_{\lambda,M}(\vy)+ 2\lambda(N-M)
\frac{\partial}{\partial\beta} \Bigr] G_{N,M}(\vx,\vy) = \Bigl(
\lambda^2 [\third(N^3-M^3) \nonu -c_0(N-M) ] + 2\lambda(N-M) [P_M(\vy)
-\half\lambda M^2  - \lambda(N+M) c_2]
\Bigr) G_{N,M}(\vx,\vy) .  \eqaend
To get from this an identity for $F_{N,M}(\vx,\vy)$ we recall Eqs.\
\Ref{G1}, \Ref{G2}, \Ref{GNM} and \Ref{FNM}, implying
\eq G_{N,M}(\vx,\vy) = \ee{\ii p (X + Y)}
F_{N,M}(\vx,\vy) \label{FG} ,\quad p = \half (N-M) \lambda . 
\eqend
Using the identities in Eq.\ \Ref{conj} this gives 
\nonueqa \Bigl[ H_{\lambda,N}(\vx) -H_{\lambda,M}(\vy) + 2\lambda(N-M)
 \frac{\partial}{\partial\beta} \Bigr] F_{N,M}(\vx,\vy) = \Bigl(
 \lambda^2 [\third(N^3-M^3) \nonu -c_0(N-M) ] + 2\lambda(N-M) [P_M(\vy) -Mp
 - \half\lambda M^2 - \lambda(N+M) c_2] \nonu - (N-M) p^2 + 2
 p[P_N(\vx)-P_M(\vy)] \Bigr) F_{N,M}(\vx,\vy) .  \nonueqaend
Inserting $p=\half \lambda(N-M)$ and using Eq.\ \Ref{later1} we see
that all terms involving $P_N(\vx)$ and $P_M(\vy)$ on the
r.h.s. cancel, and we obtain the identity in Eq.\ \Ref{rem2} with
\eqa c_{N,M} = \lambda^2 \Bigl( \third(N^3-M^3) -c_0(N-M) \Bigr)  - \lambda
(N-M)\Bigl( M\lambda(N-M) \nonu +\lambda M^2 +2 \lambda(N+M) c_2\Bigr) -
\quarter (N-M)^3 \lambda^2,  \eqaend
identical with the constant in Eq.\ \Ref{cNM}.

\bigskip

\noindent {\bf A.4 Detailed derivation of Eq.\ \Ref{trem2}.} We now
consider the functions defined in Eq.\ \Ref{tGNM}. Eqs.\ \Ref{G1},
\Ref{G2}, and \Ref{tFNM} imply
\eq \tilde G_{N,M}(\vx,\vy) = \ee{\ii (p_1X - p_2 Y) } \tilde
G_{N,M}(\vx,\vy) ,\quad p_1 = \half(\lambda N+M) ,\quad p_2=\half
(N+M/\lambda) .  \label{tGNM1}\eqend
Using Eq.\ \Ref{3} we can compute
\nonueqa 0 = \langle \Omega, [ \Phi^N_{\nu}(\vx)^* R^{N-M}
\Phi^M_{-1/\nu}(\vy), \cH^{(\nu)} ] \Omega \rangle = 
\nonu \langle [\cH^{(\nu)},  \Phi^N_{\nu}(\vx)]^* \Omega, R^{N-M}
\Phi^{-1/\nu}_M(\vy) \Omega \rangle  
\nonu 
+ \langle \Omega, \Phi_{\nu,N}(\vx)^* [ R^{N-M}, \cH^{(\nu)} ]
\Phi^{-1/\nu}_M(\vy) \Omega \rangle 
+ \langle \Omega,\Phi_{\nu,N}(\vx)^* R^{N-M}\nonu\times 
[ \lambda\cH^{(-1/\nu)} - 2(\lambda+1) \Wc^2 Q
- (\lambda^2+1/\lambda)(\third Q^3- c_0 Q),\Phi^{-1/\nu}_M(\vy)  \Omega \rangle \nonu    
= 
\Bigl[H_{\lambda,N}(\vx) + 
\lambda H_{1/\lambda,M}(\vy) - (\lambda^2+1/\lambda)(\third M^3-Mc_0)
\Bigr] \tilde G_{N,M} \nonu - \langle \Omega,\Phi_{\nu,N}(\vx)^*
[\cH^{(\nu)}, R^{N-M}] \Phi^{-1/\nu}_M(\vy)\Omega\rangle \nonu - 
2(\lambda+1) M \langle \Omega,\Phi_{\nu,N}(\vx)^* \Wc^2 R^{N-M}
\Phi^{-1/\nu}_M(\vy)\Omega\rangle .  \nonueqaend
As above, 
\nonueqa \langle\Omega,
\Phi^N_\nu(\vx)^*[\cH^{(\nu)},R^{N-M}]\Phi^M_{-1/\nu} (\vy)
\Omega\rangle = 2\lambda (N-M) \langle\Omega, \Phi^N_\nu(\vx)^* \nonu \times 
R^{N-M}\Wc^2 \Phi^M_{-1/\nu}(\vy) \Omega\rangle + \lambda^2
[\third(N^3-M^3) -c_0(N-M) ]\tilde G_{N,M}(\vx,\vy) \nonueqaend
which yields
\eqa \Bigl[H_{\lambda,N}(\vx) + \lambda H_{1/\lambda,M}(\vy)
-\lambda^2(\third N^3-Nc_0) - (\third M^3-Mc_0)/\lambda \Bigr] \tilde
G_{N,M}(\vx,\vy) \nonu = 2(\lambda N+M ) \langle\Omega,
\Phi^N_\nu(\vx)^* \Wc^2 R^{N-M}\Phi^M_{-1/\nu}(\vy) \Omega\rangle .
\eqaend
Moreover, Eq.\ \Ref{7} gives 
\eqa \langle\Omega, \Phi^N_\nu(\vx)^* \Wc^2
R^{N-M}\Phi^M_{-1/\nu}(\vy) \Omega\rangle = \Bigl( P_M(\vy) - \half
M^2/\lambda \nonu - \frac{\partial}{\partial\beta} -  
(\lambda N  + M/\lambda ) c_2 \Bigr) \tilde
G_{N,M}(\vx,\vy) , \eqaend
and with that we obtain 
\nonueqa \Bigl[H_{\lambda,N}(\vx) + \lambda H_{1/\lambda,M}(\vy) +
2(N\lambda +M) \frac{\partial}{\partial\beta} \Bigr] \tilde
G_{N,M}(\vx,\vy) = \nonu \Bigl[ \lambda^2 (\third N^3 -c_0 N) +
(\third M^3-Mc_0)/\lambda + 2(N\lambda +M) [ P_M(\vy) \nonu - \half
M^2/\lambda  -  (\lambda N  + M/\lambda ) c_2 ] \Bigr] \tilde
G_{N,M}(\vx,\vy) , \nonueqaend
or equivalently
\eqa \Bigl[H_{\lambda,N}(\vx) + \lambda H_{1/\lambda,M}(\vy) +
2(N\lambda +M) \frac{\partial}{\partial\beta} \Bigr] \tilde
F_{N,M}(\vx,\vy) \nonu = \Bigl[ \lambda^2 (\third N^3 -c_0 N) +
(\third M^3-Mc_0)/\lambda + 2(N\lambda +M) [ P_M(\vy)+ Mp_2 - \half
M^2/\lambda \nonu - (\lambda N + M/\lambda ) c_2 ] - Np_1^2 +
2p_1P_N(\vx) - \lambda M p_2^2 -2\lambda p_2 P_M(\vy) \Bigr] \tilde
F_{N,M}(\vx,\vy) , \eqaend
where we used Eqs.\ \Ref{tGNM1} and \Ref{conj}. Using Eq.\
\Ref{later1} we see that, again, all derivative terms involving
$P_N(\vx)$ and $P_M(\vy)$ cancel, and we obtain the identity in Eq.\
\Ref{trem2} with
\eqa \tilde c_{N,M} = \lambda^2 (\third N^3 -c_0 N) + (\third M^3-M
c_0)/\lambda + (N\lambda + M)[ M(N+M/\lambda) \nonu - M^2/\lambda -
2(\lambda N + M/\lambda ) c_2] - \quarter N(N\lambda +M)^2 -\quarter
\lambda M(N+M/\lambda)^2 \eqaend
identical with the constant in Eq.\ \Ref{tcNM}.

\appende

\app
\section*{Appendix B: Elementary proofs of the identities}
We define the function
\eq
\label{FNM1}
 G(\vx;\vy) = \f{ \prod_{1\leq j<k\leq N} \tet(x_{k}-x_j)^{\lambda_1} 
\prod_{1\leq J<K\leq M} 
\tet(y_{J}-y_{K})^{\lambda_2}}{\prod_{j=1}^N\prod_{K=1}^M
\tet(x_j-y_K)^{\lambda_3}} 
\eqend
and compute 
\eq W\, := \frac1G \left (\sum_{j=1}^N \frac{\partial^2}{\partial x_j^2} - A
\sum_{J=1}^M \frac{\partial^2}{\partial y_J^2} \right) G \eqend
with parameters $\lambda_1,\lambda_2,\lambda_3$ and $A$ to be
determined. We find it convenient to use in this appendix two
different kinds of indices: small letters $j,k,\ell=1,2,\ldots N$ and
capital letters $J,K,L=1,2,\ldots, M$. Obviously,
\eq \frac{\partial G}{\partial x_j}  = \left[ \sum_{j\neq k} \lambda_1
\phi(x_j-x_k) - \sum_K \lambda_3 \phi(x_j-y_K) \right] G \eqend
and
\eqa \frac1G \frac{\partial^2}{\partial x_j^2 }G = \sum_{k\neq j}
\lambda_1 \phi'(x_j-x_k) + \sum_{k,\ell\neq j}
\lambda_1^2\phi(x_j-x_k)\phi(x_j-x_\ell) - \sum_K
\lambda_3\phi'(x_j-y_K) \nonu + \sum_{K,L} \lambda_3^2
\phi(x_j-y_K)\phi(x_j-y_L) - 2\sum_{k\neq j,K} \lambda_1\lambda_3
\phi(x_j-x_k)\phi(x_j-y_K) , \eqaend
and similarly for the $y_J$-derivatives. We collect the terms in eight
different groups as follows,
\eq
W=W_1+W_2+W_3+W_4,\quad W_s = W^{(1)}_{s} - W^{(2)}_{s}
\eqend
with
\eq
\label{W11} 
W^{(1)}_{1} = \sum_{j, k\neq j} \left[
\lambda_1 \phi'(x_j-x_k) + \lambda_1^2 \phi(x_j-x_k)^2
\right] ,
\eqend
\eq \label{W12} W^{(2)}_{1} = A\sum_{J,K\neq J} \left[ \lambda_2
\phi'(y_J-y_K) + \lambda_2^2 \phi(y_J-y_K)^2 \right] , \eqend
\eq \label{W21} W^{(1)}_{2} = \sum_{j\neq k\neq \ell}
\lambda_1^2\phi(x_j-x_k)\phi(x_j-x_\ell) , \eqend
where `$\sum_{j\neq k\neq \ell}$' is short for the sum over all
$j,k,\ell$ with the constraints $j\neq k$ and $j\neq \ell$ and $k\neq
\ell$, 
\eq \label{W22} W^{(2)}_{2} = A \sum_{J\neq K\neq L}
\lambda_2^2\phi(y_J-y_K)\phi(y_J-y_L), \eqend
\eq W_{3} = (1-A) \sum_{j,K}\left[ - \lambda_3 \phi'(x_j-y_K) +
\lambda_3^2 \phi(x_j-y_K)^2 \right] \label{W3} 
\eqend
since obviously $W^{(2)}_{3} = -A W^{(1)}_{3}$, and the rest
\eqa W_4 = \sum_{j,K\neq L} \lambda_3^2 \phi(x_j-y_K) \phi(x_j-y_L) -
A \sum_{J,k\neq \ell} \lambda_3^2 \phi(y_J-x_k) \phi(y_J-x_\ell) \nonu
- \sum_{j,k\neq j,K} 2\lambda_1\lambda_3 \phi(x_j-x_k) \phi(x_j-y_K) +
A \sum_{J,K\neq J,k} 2\lambda_2\lambda_3 \phi(y_J-y_K) \phi(y_J-x_k) .
\label{W4} 
\eqaend

We insert $\phi'(x)=-V(x)$ and $\phi(x)^2 = V(x) - c_0 - 2f(x)$ (see
Eqs.\ \Ref{rel2} and \Ref{rel21}) in Eq.\ \Ref{W11} and obtain
\nonueqa W^{(1)}_{1} = \sum_{j,k\neq j} \left[
\lambda_1(\lambda_1-1) V(x_j-x_k) - \lambda_1^2 c_0 - 2 \lambda_1^2
f(x_j-x_k) \right] = \nonu
-N(N-1) \lambda_1^2 c_0 + \sum_{j<k} \left[ 2
\lambda_1(\lambda_1-1) V(x_j-x_k) - 4 \lambda_1^2 f(x_j-x_k) \right]
\nonueqaend
where we used that the functions $V$ and $f$ are even, and similarly
\nonueqa W^{(2)}_{1} = -A M(M-1) \lambda_2^2 c_0 + A \sum_{J<K} \left[ 2
\lambda_2(\lambda_2-1) V(y_J-y_K) - 4 \lambda_2^2 f(y_J-y_K) \right].
\nonueqaend
Renaming summation indices and using that $\phi$ is odd we then write
\nonueqa W^{(1)}_{2} = \sum_{j\neq k\neq \ell}
\third (-\lambda_1^2) \Bigl[
\phi(x_k-x_j)\phi(x_j-x_\ell) 
+ \phi(x_j-x_\ell)\phi(x_\ell-x_k)
+ \phi(x_\ell-x_k) \phi(x_k-x_j)  
\Bigr] . 
\nonueqaend
Inserting the identity in Eq.\ \Ref{rel2} with $x=x_k-x_j$,
$y=x_j-x_\ell$ and $z=x_\ell-x_k$ gives
\nonueqa W^{(1)}_{2} = \sum_{j\neq k\neq \ell} \third (-\lambda_1^2)
\left[ f(x_k-x_j) + f(x_j-x_\ell) + f(x_\ell-x_k) \right] 
=
-2\lambda_1^2 (N-2) \sum_{j<k} f(x_j-x_k) \nonueqaend
where we again renamed summation indices and used that $f$ is even. 
Similarly,
$$ W^{(2)}_{2} = -2A\lambda_2^2 (M-2) \sum_{J<K} f(y_J-y_K) .$$ 
Inserting Eq.\ \Ref{rel2} in Eq.\ \Ref{W3} we obtain
$$ W_3 = (1-A) \lambda_3(\lambda_3+1) \sum_{j,K}V(x_j-y_K ) -(1-A)
\lambda_3^2 \left( NM c_0 + 2 \sum_{j,K} f(x_j-y_K) \right)  .
$$ 
The first term in this expression mixes $\vx$ and $\vy$ in a
intolerable way, and in order to get a useful relation it has to
disappear. This leads to the following important restriction on
parameters,
\eq
\label{cond1} (1-A) \lambda_3(\lambda_3+1) = 0 . 
\eqend
Next we try to simplify $W_4$. We write $W_4 = W_4^{(1)} - W_4^{(2)}$ with
\nonueqa W_4^{(1)} = \sum_{j,K\neq L} \Bigl[ -\lambda_3^2 \phi(y_K -
x_j ) \phi(x_j-y_L) 
- A \lambda_2\lambda_3\phi(y_K-y_J) \phi(y_J-x_j) \nonu
- A \lambda_2\lambda_3\phi(y_J-y_K) \phi(y_K-x_j)
\Bigr] \nonueqaend
where we used that $\phi$ is odd and wrote the same term in two
different ways renaming summation indices. Similarly,
\nonueqa W_4^{(2)} = \sum_{J,k\neq \ell} \Bigl[ -A \lambda_3^2
\phi(x_k-y_J) \phi(y_J-x_\ell) 
- \lambda_1\lambda_3 \phi(x_k-x_j)
\phi(x_j-y_K) - \nonu
\lambda_1\lambda_3 \phi(x_j-x_k) \phi(x_k-y_K) \Bigr].
\nonueqaend
We now see that we can use the identity in Eq.\ \Ref{rel2} to simplify
$W_4$ provided the parameters obey the following conditions,
\eq \label{cond2} 
\lambda_3 = A \lambda_2 \quad \mbox{ and } \quad A\lambda_3 =
\lambda_1 ,  \eqend
and this is another important restriction on parameters. If and only
if this holds true we get
\nonueqa W_4^{(1)} = \sum_{j,K\neq L} (-\lambda_3^2) \Bigl[ f(y_K
-x_j) + f(y_J-y_K) + f(x_j-y_J)\Bigr] = \nonu -2(M-1) \lambda_3^2
\sum_{j,K} f(y_K -x_j) - 2N \lambda_3^2 \sum_{J<K}f(y_J-y_K),
\nonueqaend
and similarly
$$ 
W_4^{(2)} = - 2A(N-1)\lambda_3^2
\sum_{J,k} f(x_k -y_J) -
2 M  A \lambda_3^2 \sum_{j<k}f(x_j-x_k). 
$$ Assuming the conditions in Eqs.\ \Ref{cond1} and \Ref{cond2} hold
true we thus obtain
\eqa
W = 2\lambda_1(\lambda_1-1)\sum_{j<k} 
 V(x_j-x_k) - A 2
\lambda_2(\lambda_2-1)\sum_{J<K}  V(y_J-y_K) \nonu
-  \Bigl[ N(N-1) \lambda_1^2 - A M(M-1) \lambda_2^2 
+ (1-A) \lambda_3^2 NM 
\Bigr]c_0 
\nonu
+ 2( - N \lambda_1^2 + A M\lambda_3^2)  \sum_{j<k} f(x_j-x_k) +  2( 
AM \lambda_2^2 - N\lambda_3^2 ) \sum_{J<K} f(y_J-y_K) \nonu
+ 2[- (1-A) -(M-1) + A (N-1) ] \lambda_3^2  \sum_{j,K} f(x_j-y_K) ,
\eqaend
or equivalently, 
\eqa
\Bigl[ H_{\lambda_1,N}(\vx) - A H_{\lambda_2,M}(\vx) 
- C_0 
- C_1  \sum_{j<k} f(x_j-x_k) \nonu - 
C_2  \sum_{J<K} f(x_J-x_K) +  C_3 \sum_{j,K} f(x_j-y_K)\Bigr] 
G=0 \label{GG} 
\eqaend
with
\eq
C_1 = 2( N \lambda_1^2 - A M\lambda_3^2),\quad 
C_2 = 2( N\lambda_3^2 -AM\lambda_2^2) ,\quad 
C_3 = 2[A N  - M] 
\eqend
and
\eq
C_0 =  \Bigl[ N(N-1) \lambda_1^2 - A M(M-1) \lambda_2^2 
+ (1-A) \lambda_3^2 NM 
\Bigr]c_0 . 
\eqend
Using the identity in \Ref{rel3} we now compute
\eqa \frac1{G} \frac{\partial G}{\partial\beta} = \lambda_1 \sum_{j<k} [c_1 -
f(x_j-x_k) ] + \lambda_2 \sum_{J<K} [c_1 - f(y_J-y_K) ]-
\lambda_3 \sum_{j,K} [c_1 - f(x_j-y_K)] = \nonu
\frac12[N(N-1)\lambda_1 + M(M-1)\lambda_2 - 2NM\lambda_3]c_1 \nonu 
-\lambda_1 \sum_{j<k} f(y_J-y_K)  - \lambda_2 \sum_{J<K}
f(y_J-y_K) + \lambda_3 \sum_{j,K} f(x_j-y_K) . \eqaend
We thus see that we can write Eq.\ \Ref{GG} in the following form
\eq
\Bigl[ H_{\lambda_1,N}(\vx) - A H_{\lambda_2,M}(\vy) 
- \tilde C_0 + C \frac{\partial}{\partial\beta} \Bigr] 
G=0 
\eqend
provided that
\eq
\label{cond3}
C_i = C\lambda_i \quad \mbox{ for $i=1,2,3$.}
\eqend
In this case 
\eq
\tilde C_0 = C_0 + 
C \frac12[N(N-1)\lambda_1 + M(M-1)\lambda_2 - 2NM\lambda_3]c_1 . 
\eqend
Interestingly, the the conditions in \Ref{cond1}, \Ref{cond2} and
\Ref{cond3} have two non-trivial solutions. Firstly,
\eq
A=1,\quad \lambda_1= \lambda_2=\lambda_3\equiv \lambda ,
\eqend
with $C=2(N-M)\lambda$ and $\tilde C_0 = c_{N,M}$ given in Eq.\
\Ref{cNM}, and secondly,
\eq
A=-\lambda,\quad \lambda_1=\lambda, \quad \lambda_2=1/\lambda,\quad 
\lambda_3 = -1
\eqend
with $C= 2(\lambda N + M)$ and $\tilde C_0 = \tilde c_{N,M}$ in Eq.\
\Ref{tcNM}.  Obviously these two cases correspond to the identities
given in Eqs.\ \Ref{FNM}--\Ref{trem2}, and we have completed our
proof.  \QED

\appende

\noindent {\bf Acknowledgements:} I am grateful to E.\ Sklyanin and
N.\ Nekrasov for expressing their interest in the identities presented
in the paper and thus motivating me to work them out properly.  This
work was supported by the Swedish Science Research Council (VR), the
G\"oran Gustafsson Foundation, and the European grant ``ENIGMA''.


\begin{thebibliography}{99}

\bibitem[B]{B} Bernard D.: On the Wess-Zumino-Witten models on the
torus, {\em Nucl.\ Phys.\ B} {\bf 303}, 77 (1988)



\bibitem[C]{C} Calogero F.: Solution of the one-dimensional N body
problems with quadratic and/or inversely quadratic pair potentials,
{\em J.\ Math.\ Phys.\ } {\bf 12}, 419 (1971)

\bibitem[CL]{CL} Carey A.L. and Langmann E.:
Loop groups, anyons and the Calogero-Sutherland model. 
{\em Commun.\ Math.\ Phys.\ } {\bf 201} 1 (1999) 


\bibitem[EK]{EK} Etingof P.\ I.\ and Kirillov A.\ A.: On the affine
analog of Jack's and Macdonald's polynomials, {\em Duke Math.\ J.\ }
{\bf 78}, 229 (1995) [arXiv:hep-th/9403168]


\bibitem[EFK]{EFK} Etingof P.\ I., Frenkel I.\ B., and Kirillov A.\
A.: Spherical functions on affine Lie groups, {\em Duke Math.\ J.\ }
{\bf 80}, 59 (1995) [arXiv:hep-th/9407047]


\bibitem[FV]{FW1} Felder G.\ and Varchenko A.: Integral
representation of solutions of the elliptic
Knizhnik-Zamolodchikov-Bernard equations, {\em
Internat. Math. Res. Notices} {\bf 5}, 221 (1995)


\bibitem[FW]{FW2} Felder G.\ and Weiczerkowski C.: Conformal blocks
on elliptic curves and the Knizhnik-Zamolodchikov-Bernard equations,
{\em Commun.\ Math.\ Phys.\ } {\bf 176} (1996) 133
[arXiv:hep-th/9411004].


\bibitem[KMS]{KMS} Kuznetsov V.\ B., Mangazeev V.\ V. and Sklyanin E.\
K.: Q-operator and factorised separation chain for Jack's symmetric
polynomials, {\em Indag. Mathem.}, N.S., {\bf 14}, 451 (2003)
[arXiv:math.ca/0306242]

\bibitem[L1]{EL1} Langmann E.: Anyons and the elliptic
Calogero-Sutherland model, {\em Lett.\ Math.\ Phys.} {\bf 54}, 279
(2000) [{\tt math-ph/0007036}]


\bibitem[L2]{EL2} Langmann E.: Second quantization of the elliptic
Calogero-Sutherland model, {\em
Comm.\ Math.\ Phys.} {\bf 247}, 321 (2004) [{\tt math-ph/0102005}]


\bibitem[L3]{EL3} Langmann E.: Algorithms to solve the (quantum)
Sutherland model, {\em J. Math. Phys.} {\bf 42}, 4148 (2001) [{\tt
math-ph/0104039}]


\bibitem[L4]{EL4} Langmann E.: A perturbative algorithm to solve the
(quantum) elliptic Calogero-Sutherland model, {\tt math-ph/0401029}


\bibitem[OS]{OP} Olshanetsky M.A. and Perelomov A.M.: Quantum
completely integrable systems connected with semisimple Lie algebras,
Lett.\ Math.\ Phys.\ {\bf 2}, 7 (1977)


\bibitem[Sk]{Skl} Sklyanin E.\ K.: Separation Of Variables -- New
Trends, Prog.\ Theor.\ Phys.\ Suppl.\ {\bf 118}, 35 (1995)

\bibitem[Su]{Su} Sutherland B.: Exact results for a quantum many body 
problem in one-dimension. II. {\em Phys.\ Rev.} {\bf A5} 1372 (1972)


\bibitem[W]{Wen} Wen X.~G.: Chiral Luttinger liquid and the edge
excitations in the fractional Quantum Hall states, Phys.\ Rev.\ {\bf
B41}, 12838 (1990)

\bibitem[WW]{WW} Whitaker E.\ T.\ and Watson G.\ N.: {\it Course of
modern analysis, 4th edition.} Cambridge Univ.\ Press (1958)

\end{thebibliography}
\end{document}